\documentclass[review]{elsarticle}

\usepackage{hyperref}
\usepackage{times}
\usepackage{latexsym}
\usepackage{graphicx}
\usepackage{amsmath}
\usepackage{flexisym}
\usepackage{multirow}
\usepackage{array}
\usepackage{float}
\usepackage{dblfloatfix}
\usepackage{footmisc}

\journal{Journal of \LaTeX\ Templates}

\begin{document}

\begin{frontmatter}

\title{An empirical user-study of text-based nonverbal annotation systems for human-human conversations}
\tnotetext[mytitlenote]{Fully documented templates are available in the elsarticle package on \href{http://www.ctan.org/tex-archive/macros/latex/contrib/elsarticle}{CTAN}.}

%% or include affiliations in footnotes:
\author[mymainaddress]{Joshua Y. Kim}

\author[mymainaddress]{Kalina Yacef\corref{mycorrespondingauthor}}
\cortext[mycorrespondingauthor]{Corresponding author: kalina.yacef@sydney.edu.au}

\address[mymainaddress]{Computer Science Building. The University of Sydney. NSW 2006, Australia. }

\begin{abstract}
With the substantial increase in the number of online human-human conversations and the usefulness of multimodal transcripts, there is a rising need for automated multimodal transcription systems to help us better understand the conversations. In this paper, we evaluated three methods to perform multimodal transcription. They were (1) Jefferson -- an existing manual system used widely by the linguistics community, (2) MONAH -- a system that aimed to make multimodal transcripts accessible and automated, (3) MONAH+ -- a system that builds on MONAH that visualizes machine attention. Based on 104 participants responses, we found that (1) all text-based methods significantly reduced the amount of information for the human users, (2) MONAH was found to be more usable than Jefferson, (3) Jefferson's relative strength was in chronemics (pace / delay) and paralinguistics (pitch / volume) annotations, whilst MONAH's relative strength was in kinesics (body language) annotations, (4) enlarging words' font-size based on machine attention was confusing human users as loudness. These results pose considerations for researchers designing a multimodal annotation system for the masses who would like a fully-automated or human-augmented conversational analysis system.
\end{abstract}

\begin{keyword}
\texttt{elsarticle.cls}\sep \LaTeX\sep Elsevier \sep template
\MSC[2010] 00-01\sep  99-00
\end{keyword}

\end{frontmatter}

%%\linenumbers

\section{Introduction}

In recent years, there is a substantial increase in the number of human-human conversations taking place online and thereby increase the volume of conversations that could be analysed. Human-human conversations are rich in multi-modal information, nonverbal information such as those concerning tone of voice (prosody) and facial expressions could provide valuable insights into the attitudes of the speaker. A skilled communicator is therefore sensitive to nonverbal information, and there are many professions that demand strong communication skills, such as customer service and healthcare professionals (doctors, psychologists). In fact, healthcare professionals have compulsory communications courses during their undergraduate training \cite{liu2016eqclinic} -- attending to the patient's exhibited nonverbal behavior is valuable in generating a deep understanding of the patient's bio-psychosocial health \cite{beach2002body}.

To become sensitive to nonverbal information, one has to learn to recognize nonverbal information \cite{mariska2013understanding}. Whilst nonverbal information could be captured through video recordings, nonverbal information could be difficult to detect given the recording \cite{maccario2012aviation}. It takes both time and expertise to detect nuances in how one spoke, often requiring to replay the conversation segment multiple times. To make the nonverbal information obvious, conversational analysts in the linguistics field have been manually annotating such nonverbal annotations using the Jefferson Transcription System \cite{jefferson2004glossary}. In essence, the verbatim transcript is richly marked up with symbols that illustrates precisely how something is being said together with what is being said. The problem of manual transcription is that it is expensive to create such transcripts, even more so than verbatim transcriptions because the nonverbal information has to be first recognized, then subsequently weaved into the transcription at precise timings.

Since the nonverbal transcription system is expensive to produce, but is useful in revealing nonverbal information, it begs the question,  ``could nonverbal annotations be automated to some degree?" Automated annotations -- together with advances in Natural Language Processing, Computer Vision and Speech Recognition --  have the potential impact to help communication trainers or customer service managers review a large number of conversations to provide feedback to front-line workers in healthcare and customer service.

In this paper, we present the results of an experiment to evaluate three text-based  nonverbal annotation systems. The first system is the Jefferson Transcription System \cite{jefferson2004glossary}, as introduced above, it is a verbatim transcript that is richly and manually marked up with symbols representing nonverbal information. The second system is the "Multimodal Narratives for Human" (MONAH) system \cite{kim2019detecting,kim2021monah}, which is an automated system that do not use symbols but words to describe the talk turn. The third system is an add-on from the second system, it is MONAH plus words being enlarged and darkened according to the model's attention on the words (more details later).

In the experiments, the participants made use of the three systems to make sense of the emotions displayed by the speaker. We were interested in finding out which system helped participants gain a more accurate or swifter understanding of the emotions displayed, as well as the transcripts' usability.

The contribution of this work lies in:
\begin{itemize}
\item Providing an empirical assessment of multi-modal text-based annotation systems under controlled conditions.
\item Testing whether students with linguistics background have a significant difference in preference and performance from non-linguistics students.
\item Tested a method of visualizing neural network attention, through enlarging and darkening the font-face proportional to the attention weights.
\item Offering design recommendation for future nonverbal annotation systems.
\end{itemize}

\newpage

\section{Background}
Whilst text-based annotation have been used by human experts in the linguistics community to annotate nonverbal information, there have also been attempts by the computer visualization community to visualize human-human conversations. A common outcome between text-based annotation and computer visualizations is that users' awareness to nonverbal information should be increased when there is a nonverbal information extraction step followed by a nonverbal information presentation step. We first discuss research associated with the text-based manual annotations of human-human conversations, we then discuss research associated with the automated visualizations of human-human conversations.

\subsection{Text-based manual annotations of human-human conversations}

Conversation analysis researchers often use the established Jefferson Transcription system to describe the delivery of talk and other bodily conduct \cite{hepburn2013conversation}. The basic principle of the transcription is making the transcription accessible to linguistically unsophisticated readers \cite{leemann2006prosodic}. Within the qualitative research arm of conversational analysis, there are a few specialized notation systems that pertains to eye-gaze \cite{rossano2009gaze}, and gestures \cite{streeck1993gesture}. Such systems include manually supplemented drawings or photographs \cite{mondada2018multiple}, we are not concerned with such systems in this review because they are not fully text-based.

Manual text-based transcription systems are not designed to replace the video-recording, it is meant to augment it with analysis. Since the act of transcribing is an interpretative action \cite{bucholtz2000politics} and there are technological constraints in capturing the video (such as camera angle), researchers are encouraged to provide their audiovisual data along with their analyses and conclusions so that others can view the original recording and decide if they would agree \cite{jones2002research}. That said, the Jefferson Transcription system has been used in multiple professional settings. For example, police interrogations \cite{antaki2017police}, mediation \cite{alexander2020characterological}, and conversations with a member of parliament \cite{hofstetter2018getting}. A criticism to the Jeffersonian transcription is that it could be too subjective -- overly dependent on the person's interpretations. Given the usefulness of nonverbal annotations, we are taking a closer look at how the manual transcription compare against an automated approach.

\subsection{Automated visualizations of human-human conversations}

There have been attempts in visualizing the nonverbal information in metaphors such as line graphs and scatter plots. A common approach is to plot two time series (by speaker), with the Y-axis representing pitch \cite{yang2001visualizing} or volume \cite{bergstrom2007conversation}. However, because it was difficult to compare non-text-based annotations with the prevalent Jefferson transcription system which is text-based, we limited our comparison to text-based annotation systems.

\citet{kim2019review} performed a systematic review on the computer-generated visualizations built over the past two decades, and found twenty-six visualizations focused on dyadic (two-people) conversations. The authors also noted that there is a lack of multi-modal visualization weaved together with text (the subject of this paper) -- only one \cite{ghanam2008chatvis} out of the twenty-six studies visualized the keystroke aspect of human-human text chat. In a human-human face-to-face conversation, multimodal information can be group into three groups -- (1) Use of pacing of speech and length of silence (chronemics), (2) Body movements or postures (kinesic), (3) Volume, pitch, quality of voice (paralinguistics). The practice of grouping nonverbal behaviour into these three groups is prevalent in studies that study multimodal signals \cite{anvari2014determinants,poyatos2011analysis}.

The ``Multimodal Narratives for Human" (MONAH) system \cite{kim2019detecting,kim2021monah}, is an automated nonverbal annotation system. The system takes in one video per speaker and runs a pipeline of data processing steps. First, the audio is extracted and sent to an online transcription system to obtain word-level timestamped transcripts. Since there are two videos per conversation, two transcripts are obtained. Using the timestamped transcripts, talkturns are formed. The system assumes one person is speaking until the other person has interrupted, this relies on the word-level timestamps for each person's transcript. At the end of the first stage, we have the talkturns of each speaker for the entire conversation. Second, the audio and video are passed into existing pre-trained algorithms that recognize facial landmarks, tone and other features. The output of this second step are the numeric features -- both facial landmarks and tone -- accompanied with timestamp information. Third, the timestamped numeric features are weaved into the verbatim talkturns using a set of customizable rules to insert phrases that annotate nonverbal information into the verbatim transcript. 

Compared to the Jefferson transcript, a MONAH transcript appears to be easier to read because it does not involve technical symbols. The authors found that the multimodal annotations improved the performance of supervised learning \cite{kim2019detecting,kim2021monah}. However, there was no mention of whether the annotations improved the users' understanding of the conversation. In this study, we aim to compare how this automated system compare against the manual Jefferson transcription system in terms of the task of helping its users guess the emotions of a talkturn, evaluate its usability, evaluate its thoroughness in annotating various nonverbal behaviours and lastly conclude with the qualitative discussion of open-ended responses on perceived differences.

\subsubsection{Analysis of how human interpret attention from supervised deep learning models}
Deep learning models are often criticised for its lack of interpretability. Attention has attempted to offer a layer of interpretability into deep learning models. There are studies that support attention, especially in applying Deep Learning Computer Vision on healthcare images, like echocardiograms \cite{ghorbani2020deep}, radiographs \cite{lee2017fully}. Over the recent few years, researchers have introduced multiple attention layers that improve performance. With regards to interpretability, a single attention layer is easier to interpret than multiple attention layers. Take Transformers \cite{vaswani2017attention} for example, a naive approach of averaging the attention layer weights for each token would lead to the dilution of signal and would not consider the interaction between layers \cite{chefer2021transformer}.

In this paper, the attentions weights we visualized are hierarchical in nature, but both are single layer attention layers (at the word level and then at the talk turn level). Instead of producing heatmaps \cite{ghaeini2018interpreting} or highlighting the words with different colours \cite{shen2016neural}, we directly enlarge and darken the words in the visualization proportional to the attention weights to try reduce the cognitive load of the reader.

\subsection{Purpose of current study}

This study seeks to understand how people perceived different text-based multimodal annotation systems to represent videos. In this experiment, we looked at verbatim text, MONAH, MONAH+ \cite{kim2019detecting,kim2021monah} and Jefferson representations \cite{jefferson2004glossary}. This study aims to create a cheaper and faster way to translate videos in a manner that captures enough relevant information in the video to interpret the actual scenario in the conversation, by comparing the effectiveness different text-based representations of video.

\newpage
\section{Method}
\subsection{Participants}
We recruited 125 participants (63 linguistics, 62 non-linguistics) who were students from 27 universities across Australia, United States, Canada and the United Kingdom. Participants received an Amazon gift voucher for taking part in the survey, the value of the gift voucher was 10 AUD/USD/CAD/GBP, the currency corresponded to the country of their university -- Australia, United States, Canada, United Kingdom. respectively. There are no bonus provided on top of the gift voucher.

The country of participants influences culture, which in turn influences how emotions are displayed and interpreted \cite{fischer2004gender,biehl1997matsumoto}. Therefore, We controlled for the country -- the country-mix of linguistics students is identical to the country-mix of non-linguistics students (see table \ref{tab:participants_countries}). If we have more responses than required for a country, we discard the later responses. Using this method, we have discarded 21 responses. For the rest of this paper, all results are based on the remaining 104 country-balanced responses.

104 participants is a reasonably large sample for targeted recruitment -- in the review of \citet{kim2019review}, the authors compared all visualizations that investigates dyadic conversations in the past two decades, and found that the range of the number of participants ranged from 3 to 48. In another study, \citet{borgo2018information} found that for visualization evaluations using \textit{crowdsourcing}, the median number of participants is 115. In our study, however, we did not use crowdsourcing but targetted recruitment via the traditional route of advertising through the university faculty email distribution list. Therefore, we were able to ensure that the participants are currently enrolled students because the personal link to the survey was sent to the participant's university email address.

\begin{table*}[ht]
\caption{Breakdown of participants by countries. Percentage indicates the proportion of column-group contributed by the country. We also noted the number of excess responses that were discarded to balance the responses by country.}
  \centering
  \begin{tabular}{|>{\centering\arraybackslash}m{2.50cm}
                |>{\centering\arraybackslash}m{2.50cm}
                |>{\centering\arraybackslash}m{2.50cm}
                |>{\centering\arraybackslash}m{2.50cm}|}
\hline
            {\textbf{Country}}     &   {\textbf{Linguistics}} &   {\textbf{Non-Linguistics}}   &   {\textbf{Total}} \\
\hline
            {Australia}     &   {15 (29\%)} &   {15 (29\%)\newline Excess: 10}   &   {30 (29\%)} \\
\hline
            {USA}     &   {24 (46\%)\newline Excess: 1} &   {24 (46\%)}   &   {48 (46\%)} \\
\hline
            {Canada}     &   {9 (17\%)\newline Excess: 3} &   {9 (17\%)}   &   {18 (17\%)} \\
\hline
            {United Kingdom}     &   {4 (8\%)\newline Excess: 7} &   {4 (8\%)}   &   {8 (8\%)} \\
\hline
            {Total}     &   {52} &   {52}   &   {104} \\
\hline
\end{tabular}
  
  \label{tab:participants_countries}
\end{table*}

The study, including the call for participants, has been pre-approved by the ethics committee to ensure that there are no undue pressure or incentive for biased results. The project number is ``YYYY/NN"\footnote{Masked for double-blind review.}. Our call-for-recruitment process is as follow. Firstly, we identified universities who have a research presence in Linguistics research using the QS World University Ranking in 2020\footnote{https://www.topuniversities.com/university-rankings/university-subject-rankings/2020/linguistics}. We checked that the academic calendar of the university indicates that a term is in-progress before sending a request to a faculty administration staff to broadcast our recruitment email. Students would sign-up using the link provided and we would ensure that duplicates are identified via their email address and removed before sending the participant the personal link to the survey. 
We repeated this process for non-linguistics students, and reached out to the computer science and psychology faculties. From our experience, it was considerably harder to recruit for students with a linguistics background because it is a niche pool of students.

\subsection{Material}
\label{subsection:material}
Participants were exposed to five different variants of information, namely (1) video-recording, (2) verbatim transcript, (3) Jefferson, (4) MONAH, (5) MONAH+. We will introduce each of this in turn.

\begin{figure*}[h!]
    \centering
    \includegraphics[width=0.95\textwidth]{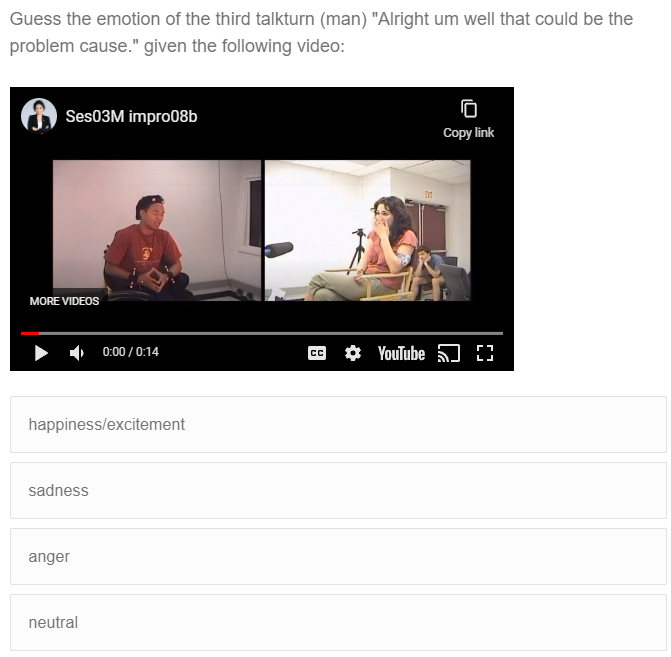}
    \caption{Participants can view the videos multiple times. Videos are hosted on YouTube.}
    \label{fig:video_representation_example}
\end{figure*} 

We begin with (1) video-recording, this is the original data supplied in the IEMOCAP dataset \cite{busso2008iemocap}. This video dataset was recorded with ten actors in five dyadic sessions (man-woman pairs), consisting of both scripted and spontaneous spoken scenarios. Each talkturn's emotion was annotated by at least three annotators and the majority label was used as the correct answer in the guess-the-emotion section of the questionnaire. This video-based dataset has been frequently used in research associated with automatic emotions recognition systems including multimodal \cite{mittal2020m3er,gu2018deep} and multitask approaches \cite{li2019improved,atmaja2020multitask}. When exposing participants to this variant (video-recording), the two videos are weaved together as one video and uploaded onto YouTube (see fig \ref{fig:video_representation_example}). The participant is able to play the video multiple times, but that would mean that participants take longer to answer the question.

Secondly, we introduce (2) verbatim transcript. This is the transcript that is returned from the Google Speech-to-text service. We chose Google Speech-to-text because it is a competitive transcription online service \cite{kim2019comparison}. The data preprocessing to yield talkturn transcript is as follows. We uploaded two audio files (one for each speaker) to the service. The service returned two result sets when the transcription is completed. The result set contains word-level timestamps -- using this timestamp information, we concatenate words together to form a talkturn until a word is spoken by the other party. When this happen, we deem that a talkturn has ended and the other party's talkturn has begun. Using this method, we bucketed timestamped words into talkturns. We also provided the previous two talkturns so that the participant can get the context to guess the emotion behind the third talkturn. Conversational context -- in terms of what was being said previously --  has been proven to be important in emotion recognition \cite{poria2017context,soujanya2017multi}.

\begin{figure*}[h]
    \centering
    \includegraphics[width=0.95\textwidth]{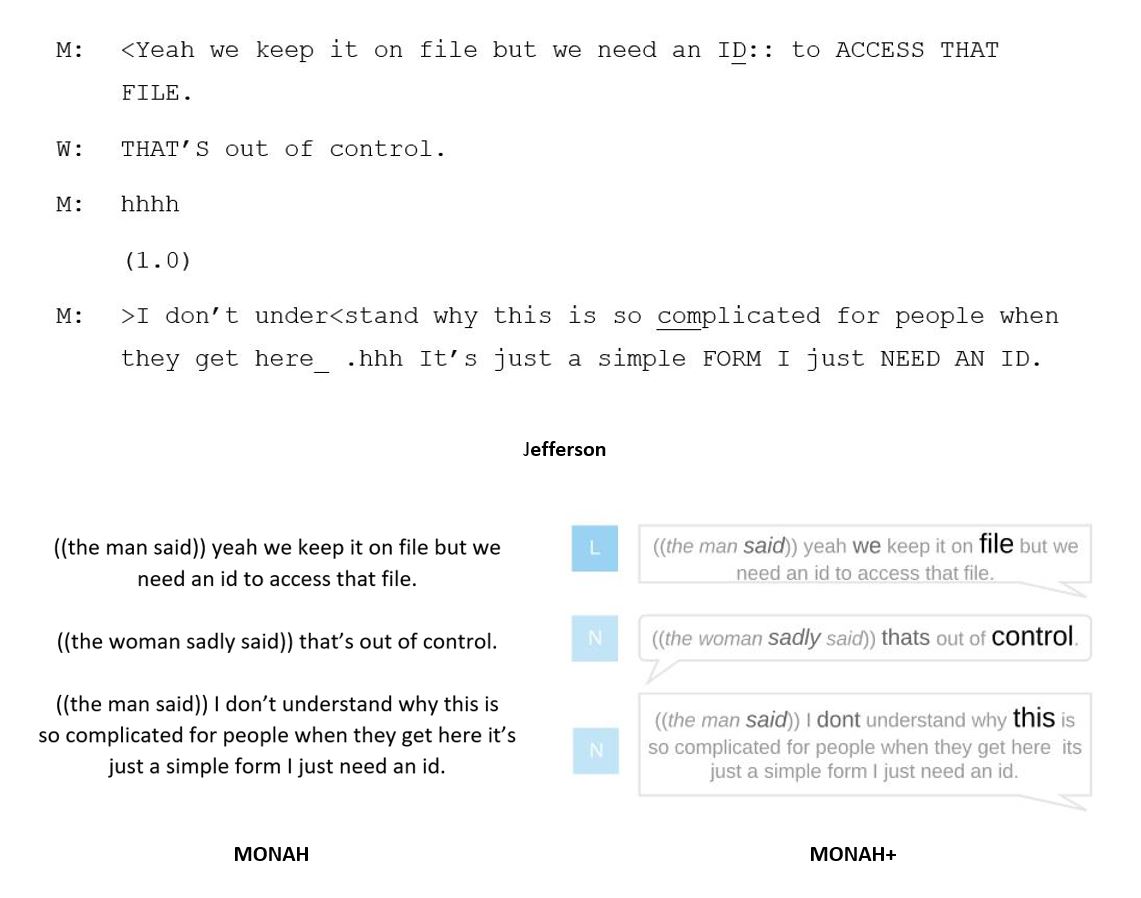}
    \caption{Sample representation using Jefferson (Top), MONAH (Lower Left), and MONAH+ (Lower Right).}
    \label{fig:JMN}
\end{figure*}

At this point, it would be helpful to present the remaining three transcription variants side by side to facilitate a visual comparison before describing Jefferson, MONAH and MONAH+ in words (see Fig. \ref{fig:JMN}).

Thirdly, we introduce (3) Jefferson transcription. We have clipped the video file to the specific three talkturns before sending them to the Jefferson Transcription Specialists at University Transcription (www.universitytranscriptions.co.uk). In Jefferson transcripts, the position of the words is indicative of the relative timings of the talkturns, therefore, we have taken screenshots from the supplied transcripts to preserve the position of the words when presenting them in the questionairre. We also introduced the glossary of the symbols to the participants as part of the training section (see section \ref{subsection:procedure}).

Fourth, we introduce (4) MONAH transcripts. The MONAH transcription system \cite{kim2019detecting,kim2021monah} uses a combination of existing computer vision models and speech analysis models to produce multimodal numeric feature tables that are timestamped. Using these feature tables, the system uses a set of predefined rules to insert phrases that describes the talkturn. \citet{kim2021monah} outlined the feature families as Demographics, Actions, Prosody, Semantics, Mimicry and History. We used the Open-Sourced System without modification.

Lastly, we introduce (5) MONAH+ transcripts. The MONAH transcript is text-based and the MONAH+ transcripts illustrates the talkturn-level and word-level attention weights from the trained model (see section \ref{subsubsection: Trainig of HAN}). The attention weights are normalized and used as the basis to enlarge and darken the font face so that its users could analyze the important segments of the conversations quickly. At a high level, we trained a Hierarchical Attention Model (HAN) \cite{yang2016hierarchical} to predict the 4-class emotion classification, and used the trained model to generate the talkturn-level and word-level attention weights. Attention has been an active area of research because of both its ability to improve performance \cite{vaswani2017attention}, and model interpretability \cite{yang2016hierarchical,xu2015show}, therefore, we wanted to test if magnifying and colouring the words using the attention weights helps human users interpret the transcripts better.

\subsubsection{Training of HAN}
\label{subsubsection: Trainig of HAN}
In this subsection, we discuss the training of HAN in greater detail. The dependent variable is a 4-class emotion label (happy, sad, neutral, angry). The proportion of the classes are as follows: happy (30\%), sad (20\%), neutral (31\%), anger (20\%). We used the HAN model \cite{yang2016hierarchical} without modification, initialising about 5 million parameters. We chose Glove  \cite{pennington2014glove} to represent the words using pre-trained vectors of 300-dimensions.

For hyper-parameter tuning, we allowed twenty trials for each configuration. During each trial, we used random search \cite{bergstra2012random} to pick a random setting for each hyper-parameters under the following boundaries. All hyperparameters are sampled on the linear scale, with the exception of learning rate which is sampled on the log-linear scale. We used a SGD optimizer with a learning rate between 0.003 to 0.010. Our batch size is between 4-20. We tuned the number of Gated Recurrent Units \cite{cho2014learning} to be between 40 and 49, whilst allowing dropout to be between 0.05-0.50 \cite{gal2016theoretically}. 

\subsection{Hypotheses}

A problem that designers of any visualization tool face is how best to visualize the content to its users so that pertinent information is communicated clearly. At the same time, an annotated transcript presents different content from the video-recording, therefore, we think that it was important to compare to two baselines -- the verbatim transcript without any nonverbal information, and the original video-recording. There are four parts to the experiment (namely, guess-the-emotion, usability, quality of annotations, and open-ended feedback), three of which are quantitative in nature which allows us to establish hypotheses and test them. The last part of the experiment is qualitative in nature. We will now discuss the hypotheses tested with regards to each of the three quantitative parts in turn. There are a total of nine hypotheses.

In the first part of the experiment (guess-the-emotion), we were interested to see if participants (or sub-groups of participants) exhibited different levels of performances in guessing-the-emotion under the different representations. We test the following hypotheses:
\begin{itemize}
\item \textbf{H1A:} \textit{The accuracy of guessing the emotions using MONAH, MONAH+ and Jefferson transcripts would outperform verbatim transcripts significantly.} The presence of multimodal annotations in all three transcripts should improve emotion recognition.
\item \textbf{H1B:} \textit{Accuracy and speed of linguistics students using the Jefferson transcripts would be significantly better than non-linguistics students using the Jefferson transcripts.} Linguistics students are trained in the Jefferson transcription system, and should be able to utilise it better than non-linguistics students, resulting in better accuracy as well as speed.
\item \textbf{H1C:} \textit{The accuracy of guessing the emotions using video-recordings would outperform all four transcripts significantly.} Text-based annotations although rich in information, should not be able to substitute for video-recordings.
\item \textbf{H1D:} \textit{For students with linguistics background, accuracy scores using Jefferson transcripts will be significantly higher than using verbatim transcripts.} Linguistics students are more familiar with the Jefferson transcripts and therefore, should be able to use the nonverbal annotations and yield better accuracy scores.
\end{itemize}

In the second part of the experiment, we asked the participants to rate the usability of the system using the SUS questionnaire. We aim to benchmark the SUS scores using the review by \citet{bangor2008empirical}. In addition, we test the following hypotheses:
\begin{itemize}
\item \textbf{H2A:} \textit{Students with linguistics background would find the Jefferson system significantly more usable than MONAH or MONAH+.} We posit that this is possible because of the training in Jefferson as part of their studies.
\item \textbf{H2B:} \textit{Students with non-linguistics background would find the MONAH or MONAH+ system significantly more usable than the Jefferson system.} We post that this is possible because the Jefferson system has many symbols which might require formal training to fully utilize it. On the other hand, the MONAH(+) system do not use technical symbols.
\end{itemize}

In the third part of the experiment, we asked the participants to rate the three aspects of nonverbal annotations. The three aspects are chronemics (use of pacing of speech and length of silence), kinesic (body movements or postures), paralinguistics (volume, pitch, and quality of voice). We test the following hypotheses:
\begin{itemize}
\item \textbf{H3A:} \textit{Jefferson would significantly outperform MONAH in chronemics score.} Both systems annotate the length of silence and pacing. MONAH annotates it using a phase ``after x milliseconds," whilst Jefferson annotates it using brackets and numbers e.g., (0.4) signifies 0.4 seconds. However, in addition to delays, Jefferson also annotates the elongation of words with colons and quick and slow speech with angle brackets. Therefore, we expect Jefferson to significantly outperform MONAH in this aspect.
\item \textbf{H3B:} \textit{MONAH would significantly outperform Jefferson in kinesic score.} MONAH has descriptions around facial expression (e.g., ``the women smiled the woman said") but not Jefferson.
\item \textbf{H3C:} \textit{Jefferson would significantly outperform MONAH in paralinguistic score.} Jefferson has a richer set of annotations around volume and pitch (e.g., ``TO ACCESS THAT FILE" signifies loud volume), whilst MONAH do not attempt to visualize volume and pitch.
\end{itemize} 

\subsection{Design}

To test the hypotheses, we have designed a sequence with four steps in the user study. At the beginning of the user study, the system randomly assigns a group number from one to five (inclusive). This group number remained constant for the participant throughout the study, we now describe what types of questions each group encountered in each step.

\subsubsection{Guess-the-emotion}
This part contained 15 questions with 3 questions for each method of representation (Verbatim Text, MONAH, MONAH+, Jefferson, and Video). Each participant was required to guess the emotion being portrayed in the third talkturn in the representation – whether the speaker is angry, neutral, happy, or sad. Fig. \ref{fig:JMN} (section \ref{subsection:material}) gives an example of how we presented the three talkturns using different representations. To be clear, a participant did not see more than one representation for each question. In our design, we controlled for question difficulty by ensuring that the set of 15 question is kept constant across groups. Each group of participant were exposed to three questions for each representation variant; also, every question had attempts on all five representation variants (see Table \ref{tab:design_guess_the_emotion}). There was no time limit but the duration of how long the participant finished guessing in each question was recorded. 

\begin{table*}[h!]
  \caption{Distribution of representation variants by the five participant groups. Row-wise, we observe that each participant group is being exposed to all five types of representation variants. Column-wise, we observe that each question has got attempts across all five types of representation variants. Therefore, we have controlled for question difficulty.}
  \centering
  \begin{tabular}{|>{\centering\arraybackslash}m{0.90cm}
                |>{\centering\arraybackslash}m{0.30cm}
                |>{\centering\arraybackslash}m{0.30cm}
                |>{\centering\arraybackslash}m{0.30cm}
                |>{\centering\arraybackslash}m{0.30cm}
                |>{\centering\arraybackslash}m{0.30cm}
                |>{\centering\arraybackslash}m{0.30cm}
                |>{\centering\arraybackslash}m{0.30cm}
                |>{\centering\arraybackslash}m{0.30cm}
                |>{\centering\arraybackslash}m{0.30cm}
                |>{\centering\arraybackslash}m{0.30cm}
                |>{\centering\arraybackslash}m{0.30cm}
                |>{\centering\arraybackslash}m{0.30cm}
                |>{\centering\arraybackslash}m{0.30cm}
                |>{\centering\arraybackslash}m{0.30cm}
                |>{\centering\arraybackslash}m{0.30cm}|}
\hline
            {\textbf{Group}}     
            & \multicolumn{15}{c|}{\textbf{Question Number}} \\
            {\textbf{No.}}     
            & {\textbf{1}} & {\textbf{2}} & {\textbf{3}}
            & {\textbf{4}} & {\textbf{5}} & {\textbf{6}}
            & {\textbf{7}} & {\textbf{8}} & {\textbf{9}}
            & {\textbf{10}} & {\textbf{11}} & {\textbf{12}}
            & {\textbf{13}} & {\textbf{14}} & {\textbf{15}}
        \\
\hline
            {1}     
            & \multicolumn{3}{c|}{Verbatim}
            & \multicolumn{3}{c|}{MONAH}
            & \multicolumn{3}{c|}{MONAH+}
            & \multicolumn{3}{c|}{Jefferson}
            & \multicolumn{3}{c|}{Video} \\
\hline
            {2}     
            & \multicolumn{3}{c|}{MONAH}
            & \multicolumn{3}{c|}{MONAH+}
            & \multicolumn{3}{c|}{Jefferson}
            & \multicolumn{3}{c|}{Video} 
            & \multicolumn{3}{c|}{Verbatim} \\
\hline
            {3}     
            & \multicolumn{3}{c|}{MONAH+}
            & \multicolumn{3}{c|}{Jefferson}
            & \multicolumn{3}{c|}{Video} 
            & \multicolumn{3}{c|}{Verbatim} 
            & \multicolumn{3}{c|}{MONAH} \\
\hline
            {4}     
            & \multicolumn{3}{c|}{Jefferson}
            & \multicolumn{3}{c|}{Video} 
            & \multicolumn{3}{c|}{Verbatim} 
            & \multicolumn{3}{c|}{MONAH} 
            & \multicolumn{3}{c|}{MONAH+} \\
\hline
            {5}     
            & \multicolumn{3}{c|}{Video} 
            & \multicolumn{3}{c|}{Verbatim} 
            & \multicolumn{3}{c|}{MONAH} 
            & \multicolumn{3}{c|}{MONAH+} 
            & \multicolumn{3}{c|}{Jefferson} \\

\hline

\end{tabular}
  
  \label{tab:design_guess_the_emotion}
\end{table*}

\subsubsection{System Usability Scale}
\begin{figure*}[h!]
    \centering
    \includegraphics[width=0.90\textwidth]{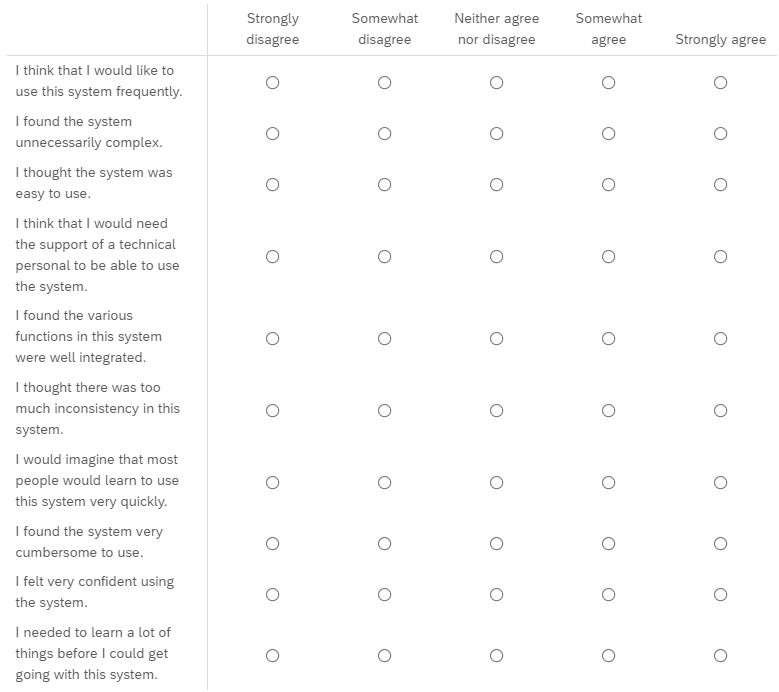}
    \caption{SUS Questions \cite{brooke1996sus} repeated for MONAH, MONAH+ and Jefferson transcripts}
    \label{fig:sus_questions}
\end{figure*}

We used the System Usability Scale (SUS) to test for usability. It was developed by \citet{brooke1996sus} as a reliable, low-cost usability scale. Over the past 15 years, this instrument has been cited more than 11,000 times and \citet{bangor2008empirical} has put together an empirical evaluation of more than 200 studies that contain SUS data. We aim to compare our data against this benchmark. In this study, all participants, regardless of the five-groups assignment, were asked to rate each of the three methods of representation usability namely, MONAH, MONAH+, and Jefferson based on the SUS questions (see Fig. \ref{fig:sus_questions}).

\subsubsection{Rate non-verbal annotation}
\begin{figure*}[h!]
    \centering
    \includegraphics[width=0.90\textwidth]{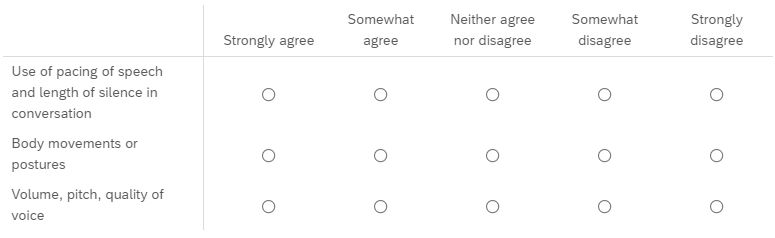}
    \caption{Rate non-verbal annotation questions presented with the 5-point Likert Scale. We insert the relevant transcript variant above the choices, and phrased the question as, "The amount of information is sufficient to interpret how the talkturn is being said."}
    \label{fig:rate_annotation_questions}
\end{figure*}

In this section, there are 15 questions, each illustrating a different conversation segment. Each of the five participant group was tasked with a set of three non-overlapping questions ($5 \times 3 = 15$). Table \ref{tab:design_rate_annotations} illustrates that each question contains 9 sub-parts, because there were 3 types of transcripts by 3 types of aspects. The participant is asked to rate the non-verbal annotations for MONAH and Jefferson. The non-verbal annotations aspects considered are the (1) Chronemics: use of pacing of speech and length of silence, (2) Kinesic: body movements or postures, and (3) Paralinguistic: volume, pitch, and quality of voice. The three representations were presented in-turn. For each representation, we asked the same set of three questions as illustrated in Fig \ref{fig:rate_annotation_questions}.

\begin{table*}[h!]
  \caption{One question contains 9 sub-parts, 3 types of transcripts (MONAH, MONAH+ and Jefferson) by 3 types of aspects (C: Chronemics; K: Kinesic; P: Paralinguistic). }
  \centering
  \begin{tabular}{|>{\centering\arraybackslash}m{2.00cm}
                |>{\centering\arraybackslash}m{0.30cm}
                |>{\centering\arraybackslash}m{0.30cm}
                |>{\centering\arraybackslash}m{0.30cm}
                |>{\centering\arraybackslash}m{0.30cm}
                |>{\centering\arraybackslash}m{0.30cm}
                |>{\centering\arraybackslash}m{0.30cm}
                |>{\centering\arraybackslash}m{0.30cm}
                |>{\centering\arraybackslash}m{0.30cm}
                |>{\centering\arraybackslash}m{0.30cm}|}
\hline

            {Transcript} 
            & \multicolumn{3}{c|}{MONAH}
            & \multicolumn{3}{c|}{MONAH+}
            & \multicolumn{3}{c|}{Jefferson}
        \\
\hline
            {Aspect} 
            & {C} & {K} & {P}
            & {C} & {K} & {P}
            & {C} & {K} & {P}
        \\
\hline
            
\end{tabular}
  
  \label{tab:design_rate_annotations}
\end{table*}

\subsubsection{Open-Ended Qualitative Questions}
Having completed the quantitative sections of the survey, we further asked participants open-ended qualitative questions to better understand their perception of the three methods of representation – MONAH, MONAH+, and Jefferson. We list the questions below before discussing them in the next paragraph:
\begin{enumerate}
\item Can you explain what you felt was difficult about guessing the emotion when not given the video?
\item What are your perceived differences between MONAH and MONAH+?
\item What are your perceived differences between MONAH / MONAH+ vs. Jefferson transcripts?
\item What are your perceived differences between MONAH / MONAH+ vs. verbatim transcripts?
\item How would you improve MONAH?
\item How would you improve MONAH+?
\item How would you improve Jefferson?
\end{enumerate}

Question 1 (``Can you explain what you felt was difficult about guessing the emotion when not given the video?") was deliberately vague to collect a range of response that could (a) help us assess what kind of information are currently missing from the transcripts, (b) help us understand the thought process of how the participants were using the existing information available in the transcript, (c) did users find providing the previous two talkturn as verbal context helpful? 

Given that MONAH and MONAH+ contained the same textual content, but MONAH+ had the additional visualization based on the attention weights, Question 2 (``What are your perceived differences between MONAH and MONAH+?") was set to help tease out the effectiveness of the attention visualization. Did the attention visualization help or hinder interpretation? Were the attention weights dependable? Perhaps the attention weights were relevant, but was the style of displaying the attention weights effective?

Question 3 (``What are your perceived differences between MONAH / MONAH+ vs. Jefferson transcripts?") narrows into the two main types of nonverbal annotation, MONAH vs. Jefferson. On one aspect, we are looking for qualitative evidence that supports or contradicts the results from the quantitative sections (i.e., accuracy from guess-the-emotion, usability scores, and rate the nonverbal annotations scores). On the other aspect, we are looking for qualitative discussions that suggests scenarios we could have tested quantitatively for future research.

Question 4 (``What are your perceived differences between MONAH / MONAH+ vs. verbatim transcripts?") attempts to answer the question that the creators of MONAH missed -- Is there any human-perceived difference between MONAH and a verbatim transcript?

Question 5 to 7 (``How would you improve MONAH / MONAH+ / Jefferson?") aimed at how each of the nonverbal annotation system -- MONAH, MONAH+ and Jefferson can further be improved, if possible. For a non-technical system like MONAH, we expected answers to be associated around how to make it even more user-friendly or more precise annotations. For Jefferson, we expected answers to be associated around making the symbols more accessible to the unsophisticated user.

\subsection{Procedure} \label{subsection:procedure}

\begin{figure*}[h!]
    \centering
    \includegraphics[width=0.85\textwidth]{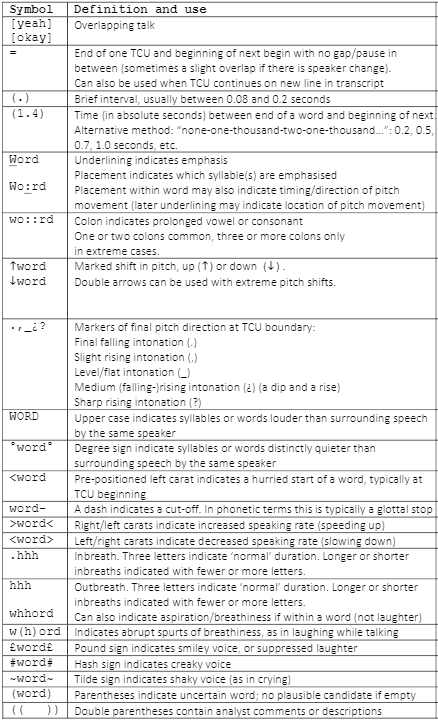}
    \caption{Participants were provided this figure at the start of session to familiarize themselves with the Jefferson Transcriptions.}
    \label{fig:jefferson_glossary}
\end{figure*}

The entire experiment lasted 40 minutes. It was conducted via the Qualtrics platform. When each participant log onto their personal links to the survey, the system assigns them randomly into one of the five groups (the reason for having the five groups are discussed in the previous section). We then have an introductory section where participants were presented two video snippets per emotion (happy, anger, sad, neutral), then participants were also presented a video snippet alongside the four types of transcripts (verbatim, Jefferson, MONAH, MONAH+). For Jefferson, the participants were also presented a glossary on what each symbol meant (see figure \ref{fig:jefferson_glossary}).

After the introductions, the participants entered the screening section. They were presented four video snippets and asked to guess the emotion for the last talkturn. This is our method of avoiding random clickers. At this point, the participant must get at least three out of four questions correct before they can proceed. Otherwise, the survey terminates with a thank-you message. Since the link to the survey is personalized, we could ensure that each participant only had one opportunity to pass this screening section. 124 out of 159 (78\%) who attempted the screening section passed it, and subsequently completed the survey. Out of the 124 responses, only 104 responses are used in this paper when we balanced the number of linguistics participants to non-linguistics participants across the four countries of study.

\newpage
\section{Results}
\subsection{Quantitative data}

We used a between-subjects analysis of variance (ANOVA) to compare scores on guessing the emotion between the five different methods of representation (Verbatim text, MONAH, MONAH+, Jefferson, and Video).  We summarized all hypotheses results in Table \ref{tab:hypotheses_results_summary}, details are discussed in the following sections.

\begin{table*}[h!]
  \caption{Summary of statistical analyses for all hypotheses. A significance level of 0.05 was used for judging the significance of difference.}
  \centering
  \begin{tabular}{|>{\centering\arraybackslash}m{9.80cm}
                |>{\centering\arraybackslash}m{1.60cm}|}
\hline
            {\textbf{Hypothesis}}     &   {\textbf{Significant?}}  \\
\hline
            {\textbf{H1A} \textit{The accuracy of guessing the emotions using MONAH, MONAH+ and Jefferson transcripts would outperform verbatim transcripts.}}     &   {No}  \\
\hline
            {\textbf{H1B} \textit{Accuracy and speed of linguistics students using the Jefferson transcripts would be better than non-linguistics students using the Jefferson transcripts.}}     &   {No for acc. and speed.}  \\
\hline
            {\textbf{H1C} \textit{The accuracy of guessing the emotions using video-recordings would outperform all four transcripts.}}     &   {Yes}  \\

\hline
            {\textbf{H1D} \textit{For students with linguistics background, accuracy scores using Jefferson transcripts would be higher than using verbatim transcripts.}}     &   {No}  \\
            
\hline
            {\textbf{H2A} \textit{Students with linguistics background would find the Jefferson system more usable than MONAH or MONAH+.}}     &   {Yes (but in the opposite direction)}  \\            
\hline
            {\textbf{H2B} \textit{Students with non-linguistics background would find the MONAH or MONAH+ system more usable than the Jefferson system.}}     &   {Yes}  \\        
            
\hline
            {\textbf{H3A} \textit{Jefferson would outperform MONAH in chronemics score.}}     &   {Yes}  \\            
    
\hline
            {\textbf{H3B} \textit{MONAH would outperform Jefferson in kinesic score.}}     &   {Yes}  \\            
\hline
            {\textbf{H3C} \textit{Jefferson would outperform MONAH in paralinguistic score.}}     &   {Yes}  \\            
\hline

\end{tabular}
  
  \label{tab:hypotheses_results_summary}
\end{table*}

\subsubsection{Guess the emotion results}

\textbf{H1A:} \textit{The accuracy of guessing the emotiong using MONAH, MONAH+ and Jefferson transcripts outperforms verbatim transcripts significantly.}

We compared the mean accuracy between verbatim transcript and Jefferson, between verbatim and MONAH, and between verbatim and MONAH+. Participants had higher mean accuracy in guessing the emotion using MONAH, MONAH+, and Jefferson transcript than verbatim transcript, but the differences were not statistically significant. Results of the analysis are summarized in Table \ref{tab:H1A_statistics_summary}. 

\begin{table*}[h!]
  \caption{T-test of accuracy scores between different text-based methods of representation vs verbatim transcript.}
  \centering
  \begin{tabular}{|>{\centering\arraybackslash}m{2.00cm}
                |>{\centering\arraybackslash}m{2.00cm}
                |>{\centering\arraybackslash}m{2.00cm}
                |>{\centering\arraybackslash}m{1.00cm}
                |>{\centering\arraybackslash}m{1.00cm}
                |>{\centering\arraybackslash}m{1.00cm}|}
\hline
            {\textbf{Reference}} & {\textbf{Method}}  & {\textbf{Accuracy}}   &   {\textbf{t}} &   {\textbf{SE}}   &   {\textbf{p}} \\
\hline
{Baseline} & {Verbatim}     &{61.72\%} &   {-} &   {-}   &   {-} \\
\hline
\multirow{3}*{Challenger}
    &{Jefferson} &{65.15\%} &{0.84} &{0.04} &{0.40} \\ \cline{2-6}
    & {MONAH}     &{67.33\%}&   {1.40} &   {0.04}   &   { 0.16} \\ \cline{2-6}
    & {MONAH+}     &{65.68\%} &   {1.00} &   {0.04}   &   {0.32} \\

\hline
\end{tabular}
  
  \label{tab:H1A_statistics_summary}
\end{table*}

\textbf{H1B:} \textit{Accuracy and speed of linguistics students using the Jefferson transcripts would be significantly better than non-linguistics students using the Jefferson transcripts.}

Linguistics participants, on the average, had higher accuracy than non-linguistics participants. Moreover, linguistics students had faster speed in guessing the emotion in the survey than non-linguistics students, on the average. However, both differences were not statistically significant. Results of the analysis are summarized in Table \ref{tab:H1B_statistics_summary}.

\begin{table*}[h!]
  \caption{\textbf{H1B} T-test of accuracy scores and speed between Linguistics and non-Linguistics participants on Jefferson transcripts.}
  \centering
  \begin{tabular}{|>{\centering\arraybackslash}m{4.00cm}
                |>{\centering\arraybackslash}m{1.66cm}
                |>{\centering\arraybackslash}m{1.40cm}
                |>{\centering\arraybackslash}m{0.80cm}
                |>{\centering\arraybackslash}m{0.60cm}
                |>{\centering\arraybackslash}m{0.60cm}|}
\hline
            {\textbf{Metrics}} &  {\textbf{Linguistics}} & {\textbf{Mean}}  &  {\textbf{t}}  &   {\textbf{SE}}   &   {\textbf{p}} \\
\hline
            \multirow{2}{*}{Accuracy} & {Yes} & {69.39\%}   &   {1.40} &   {0.06}   &   {0.17} \\
                        & {No} & {61.00\%}   &               &               &           \\
\hline
            \multirow{2}{*}{Speed (seconds to answer)} & {Yes} & {26.5}     &   {0.91}  &   {2.02}   &   {0.37} \\
                    & {No} & {24.6}      &               &               &           \\

\hline
\end{tabular}
  
  \label{tab:H1B_statistics_summary}
\end{table*}

\textbf{H1C:} \textit{The accuracy of guessing the emotions using video-recordings would outperform all four transcripts significantly.}
Statistical analysis revealed that there is enough evidence to prove that video-recording outperformed all four types of text-based transcriptions significantly. Results are summarized in Table \ref{tab:H1C_statistics_summary}.

\begin{table*}[h!]
  \caption{\textbf{H1C} T-test of accuracy scores between different text-based methods of representation vs video recording.}
  \centering
  \begin{tabular}{|>{\centering\arraybackslash}m{2.00cm}
                |>{\centering\arraybackslash}m{2.00cm}
                |>{\centering\arraybackslash}m{2.00cm}
                |>{\centering\arraybackslash}m{1.00cm}
                |>{\centering\arraybackslash}m{1.00cm}
                |>{\centering\arraybackslash}m{1.00cm}|}
\hline
            {\textbf{Reference}} & {\textbf{Method}}  & {\textbf{Accuracy}}   &   {\textbf{t}} &   {\textbf{SE}}   &   {\textbf{p}} \\
\hline
{Baseline} & {Video}     &{76.07\%} &   {-} &   {-}   &   {-} \\
\hline
\multirow{4}*{Challenger}
            & {Verbatim} &{61.72\%}    &   {-3.87} &   {0.037}   &   {0.0001} \\ \cline{2-6}
            & {Jefferson} &{65.15\%}    &   {-2.80} &   {0.039}   &   {0.0057} \\ \cline{2-6}
            & {MONAH}   &{67.33\%}  &   { -2.29} &   {0.038}   &   { 0.0231} \\ \cline{2-6}
            & {MONAH+}  &{65.68\%}   &   {-2.76} &   {0.038}   &   {0.0062} \\

\hline
\end{tabular}
  
  \label{tab:H1C_statistics_summary}
\end{table*}

\textbf{H1D:} \textit{For students with linguistics background, accuracy scores using Jefferson transcripts will be significantly higher than using verbatim transcripts.}

Although the average accuracy using the Jefferson transcripts was about 10\% points higher than the verbatim transcripts, the difference was not statistically significant due to the high variance. The results are summarized in Table \ref{tab:H1D_statistics_summary}.

\begin{table*}[h!]
  \caption{\textbf{H1D} T-test of accuracy scores between Jefferson vs Verbatim transcripts for linguistics students.}
  \centering
  \begin{tabular}{|>{\centering\arraybackslash}m{2.00cm}
                |>{\centering\arraybackslash}m{2.00cm}
                |>{\centering\arraybackslash}m{2.00cm}
                |>{\centering\arraybackslash}m{1.00cm}
                |>{\centering\arraybackslash}m{1.00cm}
                |>{\centering\arraybackslash}m{1.00cm}|}
\hline
            {\textbf{Reference}} & {\textbf{Method}}  & {\textbf{Accuracy}}   &   {\textbf{t}} &   {\textbf{SE}}   &   {\textbf{p}} \\
\hline
{Baseline} & {Verbatim}     &{59.67\%} &   {-} &   {-}   &   {-} \\
\hline
{Challenger} & {Jefferson}     &{69.39\%} &   {1.80} &   {0.054}   &   {0.0754} \\
\hline
\end{tabular}
  
  \label{tab:H1D_statistics_summary}
\end{table*}

\newpage
\subsubsection{Usability results}

Before discussing the two individual hypothesis, we summarized the average SUS score by method and background in Table  \ref{tab:sus_score_summary}.
\begin{table*}[h]
  \caption{Average SUS score by method. Standard deviation in brackets.}
  \centering
  \begin{tabular}{|>{\centering\arraybackslash}m{2.50cm}
                |>{\centering\arraybackslash}m{2.50cm}
                |>{\centering\arraybackslash}m{2.50cm}
                |>{\centering\arraybackslash}m{2.50cm}|}
\hline
            {\textbf{Method}}     &   {\textbf{Linguistics}} &   {\textbf{Non-Linguistics}}   &   {\textbf{Total}} \\
\hline
            {Jefferson}     &   {44.1 (23.8)} &   {39.2 (21.7)}   &   {41.6 (22.8)} \\
\hline
            {MONAH}     &   {69.3 (18.5)} &   {66.9 (19.8)}   &   {68.1 (19.1)} \\
\hline
            {MONAH+}     &   {57.6 (24.1)} &   {59.5 (21.3)}   &   {58.6 (22.6)} \\

\hline
\end{tabular}
  
  \label{tab:sus_score_summary}
\end{table*}

\textbf{H2A:} \textit{Students with linguistics background would find the Jefferson system significantly more usable than MONAH or MONAH+.}

To our surprise, statistical analysis revealed that students with linguistics background found MONAH and MONAH+ significantly more usable than Jefferson. The results were in the opposite effect as stated in the hypothesis. Statistical results are summarized in Table \ref{tab:H2A_statistics_summary}.

\begin{table*}[h!]
  \caption{T-test of SUS scores of linguistics students between MONAH and MONAH+ vs Jefferson transcript.}
  \centering
  \begin{tabular}{|>{\centering\arraybackslash}m{2.00cm}
                |>{\centering\arraybackslash}m{1.90cm}
                |>{\centering\arraybackslash}m{1.90cm}
                |>{\centering\arraybackslash}m{0.90cm}
                |>{\centering\arraybackslash}m{1.00cm}
                |>{\centering\arraybackslash}m{1.40cm}|}
\hline
            {\textbf{Reference}} & {\textbf{Method}}  & {\textbf{SUS Score}}   &   {\textbf{t}} &   {\textbf{SE}}   &   {\textbf{p}} \\
\hline
{Baseline} & {Jefferson}     &{44.07} &   {-} &   {-}   &   {-} \\
\hline
\multirow{2}*{Challenger}
            & {MONAH}   &{69.31 }  &   { 5.99} &   {4.22}   &   {$\leq 0.0001$} \\ \cline{2-6}
            & {MONAH+}  &{57.60 }   &   {2.85} &   {4.74}   &   {0.0053} \\

\hline
\end{tabular}
  
  \label{tab:H2A_statistics_summary}
\end{table*}

\textbf{H2B:} \textit{Students with non-linguistics background would find the MONAH or MONAH+ system significantly more usable than the Jefferson system.}

On the flip-side, we investigated our second hypothesis on SUS scores, whether non-linguistics participants find MONAH and MONAH+ more usable than Jefferson. Statistically speaking, there is enough evidence that proves non-linguistics participants find MONAH and MONAH+ significantly more usable than Jefferson. Values are summarized in Table \ref{tab:H2B_statistics_summary}.

\begin{table*}[h!]
  \caption{T-test of SUS scores of non-linguistics students between MONAH and MONAH+ vs Jefferson transcript.}
  \centering
  \begin{tabular}{|>{\centering\arraybackslash}m{2.00cm}
                |>{\centering\arraybackslash}m{1.90cm}
                |>{\centering\arraybackslash}m{1.90cm}
                |>{\centering\arraybackslash}m{0.90cm}
                |>{\centering\arraybackslash}m{1.00cm}
                |>{\centering\arraybackslash}m{1.40cm}|}
\hline
            {\textbf{Reference}} & {\textbf{Method}}  & {\textbf{SUS Score}}   &   {\textbf{t}} &   {\textbf{SE}}   &   {\textbf{p}} \\
\hline
{Baseline} & {Jefferson}     &{39.22} &   {-} &   {-}   &   {-} \\
\hline
\multirow{2}*{Challenger}
            & {MONAH}   &{66.91 }  &   { 6.73} &   {4.12}   &   {$\leq 0.0001$} \\ \cline{2-6}
            & {MONAH+}  &{59.51 }   &   {4.76} &   {4.26}   &   {$\leq 0.0001$} \\

\hline
\end{tabular}
  
  \label{tab:H2B_statistics_summary}
\end{table*}

\newpage
\subsubsection{Results from aspects of non-verbal annotations}
Since the three hypotheses are similar and only testing three different dimensions, we summarize the results of all three hypotheses in one table (see Table \ref{tab:H3_statistics_summary}). We found significant differences in all three hypotheses. That is, Jefferson significantly outperformed MONAH in chronemics (\textbf{H3A}: use of pacing of speech and length of silence) and paralinguistic (\textbf{H3C}: volume, pitch, and quality of voice). Also, MONAH significantly outperformed Jefferson in kinesic (\textbf{H3B}: body movements or postures).

\begin{table*}[h!]
  \caption{T-test of scores from rating the three aspects of non-verbal annotation between MONAH vs Jefferson.}
  \centering
  \begin{tabular}{|>{\centering\arraybackslash}m{3.40cm}
                |>{\centering\arraybackslash}m{1.66cm}
                |>{\centering\arraybackslash}m{1.30cm}
                |>{\centering\arraybackslash}m{0.70cm}
                |>{\centering\arraybackslash}m{0.60cm}
                |>{\centering\arraybackslash}m{1.40cm}|}
\hline
            {\textbf{Aspect (Hypothesis)}} &  {\textbf{Method}} & {\textbf{Mean}}  &  {\textbf{t}}  &   {\textbf{SE}}   &   {\textbf{p}} \\
\hline
            \multirow{2}{*}{Chronemics (\textbf{H3A})} & {Jefferson} & {4.12}   &   {8.2} &   {0.10}   &   {$\leq 0.0001$} \\
            & {MONAH} & {3.28}   &               &               &           \\
\hline
            \multirow{2}{*}{Kinesics (\textbf{H3A})} & {Jefferson} & {2.66}   &   {2.6} &   {0.11}   &   {0.0089} \\
            & {MONAH} & {2.96}   &               &               &           \\
            
\hline
            \multirow{2}{*}{Paralinguistic (\textbf{H3C})} & {Jefferson} & {3.92}   &   {9.7} &   {0.10}   &   {$\leq 0.0001$} \\
            & {MONAH} & {2.91}   &               &               &           \\

\hline
\end{tabular}
  
  \label{tab:H3_statistics_summary}
\end{table*}

\newpage
\subsection{Qualitative data}
As an overview, we tabulate the length of response for each open-ended question in Table \ref{tab:word_count_qualitative}. On average, the question with the longest response (23 words) were the responses to question 3 (``what are your perceived differences between MONAH(+) vs. Jefferson transcripts?") by linguistics students. On the other hand, the question with the shortest response (9 words) were the responses to question 5 (``how would you improve MONAH?") by non-linguistics students.

\begin{table*}[h!]
  \caption{Average word count for each question by student population. Standard deviation is presented in brackets.}
  \centering
  \begin{tabular}{|>{\centering\arraybackslash}m{2.50cm}
                |>{\centering\arraybackslash}m{2.50cm}
                |>{\centering\arraybackslash}m{2.50cm}
                |>{\centering\arraybackslash}m{2.50cm}|}
\hline
            {\textbf{Question}}     &   {\textbf{Linguistics}} &   {\textbf{Non-Linguistics}}   &   {\textbf{Overall}} \\
\hline      {1}     &   {19 (18)} &   {14 (11)}   &   {17 (15)} \\
\hline      {2}     &   {17 (10)} &   {15 (13)}   &   {16 (12)} \\
\hline      {3}     &   {23 (20)} &   {18 (15)}   &   {21 (17)} \\
\hline      {4}     &   {16 (12)} &   {13 (11)}   &   {14 (12)} \\
\hline      {5}     &   {15 (15)} &   {9 (11)}   &   {12 (14)} \\
\hline      {6}     &   {15 (15)} &   {11 (12)}   &   {13 (13)} \\
\hline      {7}     &   {19 (17)} &   {12 (14)}   &   {15 (15)} \\

\hline
\end{tabular}
  \label{tab:word_count_qualitative}
\end{table*}

\subsubsection{Can you explain what you felt was difficult about guessing the emotion when not given the video?}

The first open-ended question asked the participants what they felt was difficult on guessing the emotion when the video was not provided. Many from the participants pointed out that ``lack of facial expression and body language makes it difficult to guess". Many participants highlighted as well the relevance of tone or intonation, pitch, volume of voice, and timing in understanding the emotion of the speaker.  Most of the participants highlighted the difficulty with the lack of facial expression and body language, lack of information on tone, volume, pitch and pauses of the speaker. Some of them mentioned about their confusion on technicalities like not having a clear definition of each emotion, the validity of the question whether the emotion being asked was about what the person ``feel", what the person ``convey", or what emotion they ``would conventionally be", and the very subtle difference between angry vs. sad vs. neutral. Other participants commented on the discrepancies from gestures and tones, e.g. ``just because a customer service representative is smiling does not make them happy", and on how they felt that some dialogues seem artificial. Others commented on the challenge of having incomplete, very little to no context about the emotion being provided only with verbatim transcript. On the other extreme, albeit rarely, a few participants found no difficulty not being given the video at all.

\subsubsection{What are your perceived differences between MONAH and MONAH+?}
The second question asked participants about what differences they perceived between MONAH and MONAH+. Many participants noted that certain phrases or emotions were emphasized in MONAH+ by being visual in presenting it. Others correctly mentioned that MONAH seem to be the basis for MONAH+. 

Proponents of MONAH+ noted that ``MONAH+ gives a bit more information by giving more importance to certain words." Detractors of MONAH+ suggested that MONAH is much simpler and straightforward hence the easier to use and interpret. In addition, a few participants were confused on the use of enlarged fonts, and mistakenly commented that the fonts indicates the volume or intonation of certain words in conversations.

The following are a few verbatim comments from the respondents for this question:

\begin{itemize}
    \item ``MONAH+ visually relays more information about relative pitch and volume."
    \item ``MONAH+ does a much better job of showing emphasis and highlighting the key words that it is using to make judgements"
    \item ``I liked the bolded text in MONAH+. It emphasised things that I initially thought in MONAH or didn't and changed my mind. It helped figure things out. Also aesthetically it kind of looked like a chat."
    \item ``Intonation is used in MONAH+ which allows the reader to guess pace and emotion more effectively."
    \item ``MONAH+ adds unnecessary information : you don't need to be told what words the machine thinks are important, you can get that from the context. Some of MONAH+'s choices didn't seem relevant, either. Just MONAH is fine, because it clearly indicates silences, rythm and some non-verbal cues."
    \item ``MONAH+ seemed to highlight words that were not helpful at all. I find the speech bubbles helpful but having transcriber's comments in them makes it even more confusing."
    \item ``MONAH+ frequently left me feeling confused, as occasionally a word was enlarged that I felt didn't contribute to the emotional content of the speech"
    \item ``MONAH+ was harder to use because I'm used to larger or bold words indicating tonal emphasis, rather than overall relevance."
\end{itemize}

\subsubsection{What are your perceived differences between MONAH / MONAH+ vs. Jefferson transcripts?}
The third question asked participants about the differences they perceived between Jefferson and MONAH / MONAH+. Many participants spoke about the complexity of comprehending Jefferson. Participants commented that Jefferson is ``very technical and gives a lot of information, but is not easily readable". Others found Jefferson confusing since memorizing ``various symbols were cumbersome and overwhelming". Below are some of the comments about Jefferson on being more precise:

\begin{itemize}
    \item ``Jefferson seems more precise and less confusing."
    \item ``Jefferson is extremely precise in terms of showing exactly and more objectively what was said and how it was said, clearly shows pitch changes and timing..."
    \item ``Jefferson struck me as much more detailed (maybe too detailed) than the MONAHs. It captures things like speech rate and pauses very well, but requires a lot of learning."
\end{itemize}

Others highlighted that Jefferson ``communicated more silences and speech patterns", and provided ``length of pause between conversational turns or whether someone inhaled and exhaled as a way to express emotion" which they find helpful in understanding the emotion of the speaker.

Participants pointed out that MONAH / MONAH+ seemed simpler and easier to use than Jefferson. The former contains less details but is more intuitive. Others elaborated how MONAH / MONAH+ is easier to understand that they don't need to check the documentation to use it. On the other hand, they described Jefferson to be more precise, more detailed and utilises unique symbols but is too complicated. They mentioned that Jefferson was more scientific and technical that it ``requires a lot more knowledge of the symbols, and generally much harder to read and interpret". Although they think that Jefferson is more accurate in transcribing how the conversation actually happened, it was said that it is ``cumbersome and difficult" to detail about the pauses and sounds, and that it is ``more like a device useful for linguistics and other fields analysing speech". A few participants commented that MONAH / MONAH+ focuses more on the facial expressions, while Jefferson focuses more on verbals - tone, pitch, pauses, etc.

\subsubsection{What are your perceived differences between MONAH / MONAH+ vs. verbatim transcripts?}
The fourth question asked the participants what differences they see between MONAH / MONAH+ and verbatim transcripts. Participants pointed out the difficulty on picking up the nonverbal cues of the speaker when presented only with verbatim transcripts. Here are a few comments from the participants:

\begin{itemize}
    \item ``Verbatim lacks a ton of information about tone, silences, rhythm, body language, etc."
    \item ``The verbatim transcripts don't give information about the context."
    \item ``Verbatim transcripts contain no information to help guess emotions"
\end{itemize}

Most responses were focused on highlighting MONAH / MONAH+ features. One of the most mentioned features is that MONAH / MONAH+ provides more information about emotion with adverbs like ``happily", ``sadly", or phrases like ``she smiled". Participants also mentioned about the significance of pauses, pacing, body movement, gestures, and posture used in MONAH / MONAH+ in giving more context to the speaker's emotion. Other respondents explained that MONAH / MONAH+ ``described significant facial expressions, tone of voice, and length of pauses." and such information made ``it easier to imagine [themselves] in the scenario even without the video". However, some participants commented that the added descriptions left them being confused at times, and that these descriptions ``may introduce biases or take unnecessary time to read".

Finally, other participants shared that MONAH / MONAH+ was ``much better [than verbatim transcript] with little learning [experience] to understand what they [the speakers] are doing", or in simpler words, they provided easier access to the nonverbal cues of the conversation.

\subsubsection{How would you improve MONAH?}
The fifth question gave participants the chance to suggest any further improvements they can think of having been exposed to MONAH transcripts themselves. Some find MONAH easy to understand and not felt the need to improve it. 

We first summarize the suggestions to (1) change the presentation of current annotations, and (2) increase the range of non-verbal annotations being captured.

Suggestions to change the presentation of current annotations included (a) the simplification of delay timing such as  ``one hundred milliseconds"; (b) visibly separate nonverbal annotations from verbal content; (c) within verbal content, visibly separate pause-fillers (e.g. `um') from content words; (d) using capital letters for loud and italics for quiet words, or similarly, using font size.

Suggestions to increase the range of non-verbal annotations being captured included (a) marking pitch changes; (b) marking gaze (e.g., ``looking down"); (c) marking posture (e.g., ``lean back", ``arms crossed").

\subsubsection{How would you improve MONAH+?}
The sixth question asked participants what they can further improve on MONAH+. One of the most commented at is the font sizing mechanism. Participants thought that they are ``arbitrary and distracting" that it seems to ``disrupt the flow of reading". Suggestions to change how attention is visualized included the following:

\begin{itemize}
    \item ``Instead of varying text size, have an annotation line on top of each content line (like a music sheet)"
    \item ``Some system of showing what emotion the emphasized words had--maybe color coding on a gradient"
    \item ``Highlight/bold not words the algorithm thinks is necessary (it reads as awkward and clunky) but rather the emphasised words and the louder ones"
    \item ``Rather than increasing the size of the text, use italics or bold. Increasing the size just made me emphasise words that weren't emphasised in the videos."
\end{itemize}
    
Apart from attention visualization, other suggestions were applicable to both MONAH and MONAH+, (a) Use Arabic numerals rather than words to mark pauses; (b) employ pitch arrows from Jefferson to indicate pitch; (c) extract more nonverbal cues, such as sighs and body posture; (d) more granular annotations -- marking the start and end of a laughing voice within the talkturn.

\subsubsection{How would you improve Jefferson?}
The last open-ended question asked participants how they want to improve Jefferson.  There were three main areas of improvements, (1) reference material; (2) grouping or dropping symbols; (3) presentation of information.

The first area of improvement was the easy access to the Jefferson glossary whilst interacting with the Jeffersonian transcripts. Participants remarked that it would take a few days or weeks of frequent usage to remember the symbols, until then, an easy access to the glossary either through a cheat sheet or a second monitor on-screen would be helpful.

The second area of improvement was the grouping or dropping of symbols. Related to the previous point, the vast vocabulary of Jefferson symbols prompted participants to suggest to limit the diversity through grouping symbols of together or dropping infrequently used symbols. 

Lastly, the third area of improvement was about the presentation of information. Suggestions included (1) using a range of symbols that are more intuitive. The range of textual, uni-code characters have expanded to be able to represent emoticons, perhaps there is a more efficient and intuitive legend given the expanded range of textual characters; (2) incorporate other system to capture embodied actions to complement the Jefferson system. We believe that the participants are referring to the multimodal annotation systems that were out-of-scope in this study  \cite{rossano2009gaze,streeck1993gesture,mondada2018multiple}; (3) breathing cues could be due to microphone placement and non-content. We believe this is well-explored as the act of transcribing is a subjective, interpretive act; (4) using the darkened or larger fonts from MONAH+ to reduce the diversity of symbols.

There were also many responses that indicated there was not much that could be improved upon, and there were two main reasons mentioned. Firstly, some found Jefferson as already a great system, perfect, comprehensive, and transparent, thus, nothing more to improve. Secondly, some found Jefferson overly complex and felt that they did not understand the system enough to comment about its improvement.

\newpage
\section{Discussion}

In this section, we follow the structure in the results section to discuss results from the (1) guess-the-emotion, (2) usability, (3) aspects of non-verbal annotations sections. Where applicable, we would also discuss the applicable qualitative suggestions. We finish this section with (4) the discussion on attention visualization.

\subsection{Discussion of guess-the-emotions results}

Whilst previous research found that MONAH helped machines improve supervised learning significantly (\cite{kim2019detecting,kim2021monah}), in this paper where nonverbal annotations were presented to human users, none of the nonverbal annotations (MONAH, Jefferson) improved guess-the-emotion significantly over the verbatim transcripts (\textbf{H1A}). Video, on the other hand, significantly out-performed all four text-based transcriptions significantly (\textbf{H1C}). This pair of findings suggests that whilst nonverbal annotations help compress the multimodal information into text reducing the memory or disk space requirements to store the nonverbal information, such compression leads to significant information loss to human users.

For Jefferson transcripts, we did not find that linguistics participants were able to make significantly better use of the Jefferson transcript than non-linguistics participants (\textbf{H1B}). We defined ``better use" as higher accuracy and/or faster speed, neither of which were statistically significant. From the qualitative feedback, both linguistics and non-linguistics participants raised the issue of complexity with Jefferson transcripts. This suggests that it takes more regular and frequent usage than just university coursework exposure to become comfortable or proficient with the Jefferson transcripts.

\subsection{Discussion of usability results}
Across all respondents, MONAH got the highest average score of 68.1, followed by MONAH+ at 58.6 and Jefferson at 41.6 (Table \ref{tab:sus_score_summary}). Using the benchmark set by \citet{bangor2008empirical}. This means that MONAH is on the ``high marginal acceptable" level, whilst MONAH+ is on the ``low marginal acceptable" level and Jefferson is on the ``not acceptable" level.

For students with linguistics background, we expected students with linguistics background to find the Jefferson system significantly more usable than MONAH or MONAH+ (\textbf{H2A}), because of the familiarity of the transcription system as the transcription system is widely used and taught in linguistics faculties \cite{lester2018applied}. To our surprise, we found the opposite effect to be significant -- students with linguistics background found the MONAH or MONAH+ significantly more usable than Jefferson.

As for the participants without a linguistics background, we found that they found MONAH or MONAH+ significantly more usable than Jefferson (\textbf{H2B}). Together with \textbf{H1B}, these three findings pose a consideration for researchers seeking to build a conversational analysis product for the masses or for the people with disability -- whilst Jefferson is designed for the trained analyst, it is not user-friendly to the vast majority of the population. Non-Linguistics students gave an average score of 39.2 to the Jefferson Transcription system, which is on the ``not acceptable" level.

\subsection{Discussion of results from aspects non-verbal annotations}

Jefferson significantly outperformed MONAH in both chronemics (length of silence and pacing, \textbf{H3A}) and paralinguistics (volume and pitch, \textbf{H3C}). We first discuss the feasibility of automatic extraction before discussing the method of presentation of non-verbal information.

Regarding the feasibility of automatic extraction, for the pacing of syllable pronounciation, because of MONAH's reliance on Google Speech to text, the returned timings were at word-level instead of the required syllable-level. Commercial systems typically return timings at word-level or sentence-level and it would take a specialized speech recognition system to return syllable-level information \cite{dash2018speech}. As for the volume and pitch, granular timestamped information could be readily extracted through open-source packages like OpenSmile \cite{eyben2010opensmile}.

Regarding the method of presentation, the differential annotations of word elongations with colons and speed of words articulation with angle brackets is the likely contributor for the better performance. However, since non-linguistics users find the symbols counter-intuitive, perhaps researchers could get inspiration from Read N' Karaoke where the pitch and volume of each word is lightly drawn out in the background as a sparkline behind the verbatim transcript \cite{patel2011readn}, or from \citet{ahmed2020data} who visualized the average, maximum and minimum pitch for transgender voice training. For the purpose of conversational analysis, it is likely that the analyst would need detailed pitch and volume information at the word-level (if not syllable-level) hence we posit that Read N' Karaoke is likely to work better. There was a participant suggestion to present pitch information on a musical notation -- it would take a future study to understand what different groups of users prefer, but we posit that musical notation would be less preferred for people who do not have musical training.

On the other hand, MONAH significantly outperformed Jefferson in Kinesics (body movements or postures, \textbf{H3B}). We refer the reader to \citet{kim2021monah} for the method of automatic extraction of body movement and posture. Some participants made suggestions of additional automatic extraction of body movement such as ``(look down)". Using a combination of gaze / pose extraction and timestamps, such information can be computed with the help of open-source libraries like OpenFace (gaze) \cite{baltrusaitis2018openface} and OpenPose (pose) \cite{cao2019openpose}. As for manual methods, there were other systems that included manually supplemented drawings or photographs \cite{rossano2009gaze,streeck1993gesture,mondada2018multiple}, these systems were out-of-scope of this study because they are not text-based.

\subsection{Discussion on attention visualization}
Darkening fonts and enlarging words as a method of visualizing attention was a bad design. From Table \ref{tab:H1C_statistics_summary}, we saw that MONAH+ (65.7\%) did not improve accuracy in guess-the-emotion relative to MONAH (67.3\%). In addition, we saw that MONAH+ received lower usability scores than MONAH across all student sub-populations (Table \ref{tab:sus_score_summary}).

The style of visualization (enlarging font and darkened colour) disrupted the flow of reading. Some participants are used to associating larger font with louder volume and therefore found it cognitively hard to associate larger font with higher attention. Indeed, some researchers have used a large font to indicate louder volume \cite{king2004webbie}. In future research, researchers could investigate whether darkening font without enlarging words would be a better design for attention visualization.

\section{Limitations}

In the videos that we used in the study, actors may not reflect real world scenarios as these are simulated videos only wherein actors follow and re-enact a certain script. This limitation may compromise the realism of the experiment.

Although we have controlled for the country of universities of the participants, we could not control for factors like gender and age which might also have an impact in emotions recognition and conversation analysis abilities \cite{abbruzzese2019age,olderbak2019sex}. Nevertheless, we posit that the large sample size of 104 participants help alleviate such issues.

To limit the number of questions and hence the length of the survey, we elected to use only one dataset (IEMOCAP). The type of conversations is therefore limited. The scenarios in the dataset is mostly related to everyday life, like those between friends and families, as well as a couple of scenarios between a customer service agent and a customer. Specifically, highly professional settings are not present, for example doctor-patient consultations, police interrogations or court hearings. Employees in professional settings typically mask emotions -- both positive and negative \cite{kramer2002communication} -- displaying a different range of emotions.

We assumed that students from the linguistics faculty would be more familiar with the Jefferson transcription system. Whilst we are able to verify that they are indeed current students from a linguistics faculty, the limitation is that there might be a proportion of linguistics students who have not been taught the Jefferson transcription system when doing this survey.

\newpage
\section{Conclusion}
Human-computer interaction technologies can help us manage the quickly growing volume of online human-human conversations. A prerequisite of becoming skilled in understanding human-human conversations is associated with the receiver's ability to detect nuanced non-verbal cues. Multimodal annotation software can now help us surface these non-verbal cues so that users can become aware of them. The purpose of this study was to provide empirical evidence for an application to provide multimodal annotations that are simple and easy to understand. We also discussed areas of improvements to both the automated and manual systems. These findings would serve as useful reference for future research that specialize in both automatic or human-augmented conversation analysis.

\newpage
\bibliography{mybibfile}

\end{document}